\documentclass[%
 reprint,
superscriptaddress,
 amsmath,amssymb,
 aps,
 prc,
floatfix,
]{revtex4-1}

\usepackage{graphicx}
\usepackage{dcolumn}
\usepackage{bm}
\usepackage{amssymb}
\usepackage[T1]{fontenc}

\begin{document}

\newcommand{\bbzn}[0]{$0 \nu \beta \beta$ }
\newcommand{\bbtn}[0]{$2 \nu \beta \beta$ }
\renewcommand{\arraystretch}{1.2}

\preprint{APS/123-QED}

\title{Search for \bbtn decay of $^{136}$Xe to the 0$_1^+$ excited state of $^{136}$Ba with the EXO-200 liquid xenon detector}

\newcommand{\Alabama}{\affiliation{Department of Physics and Astronomy, University of Alabama, Tuscaloosa, Alabama 35487, USA}}
\newcommand{\Alberta}{\affiliation{University of Alberta, Edmonton, Alberta, Canada}}
\newcommand{\Bern}{\affiliation{LHEP, Albert Einstein Center, University of Bern, Bern, Switzerland}}
\newcommand{\CALTECH}{\affiliation{Kellogg Lab, Caltech, Pasadena, California 91125, USA}}
\newcommand{\Carleton}{\affiliation{Physics Department, Carleton University, Ottawa, Ontario K1S 5B6, Canada}}
\newcommand{\CSU}{\affiliation{Physics Department, Colorado State University, Fort Collins, Colorado 80523, USA}}
\newcommand{\Drexel}{\affiliation{Department of Physics, Drexel University, Philadelphia, Pennsylvania 19104, USA}}
\newcommand{\Duke}{\affiliation{Department of Physics, Duke University, and Triangle Universities Nuclear Laboratory (TUNL), Durham, North Carolina 27708, USA}}
\newcommand{\IBS}{\affiliation{IBS Center for Underground Physics, Daejeon, Korea}}
\newcommand{\IHEP}{\affiliation{Institute of High Energy Physics, Beijing, China}}
\newcommand{\Illinois}{\affiliation{Physics Department, University of Illinois, Urbana-Champaign, Illinois 61801, USA}}
\newcommand{\Indiana}{\affiliation{Physics Department and CEEM, Indiana University, Bloomington, Indiana 47405, USA}}
\newcommand{\ITEP}{\affiliation{Institute for Theoretical and Experimental Physics, Moscow, Russia}}
\newcommand{\Laurentian}{\affiliation{Department of Physics, Laurentian University, Sudbury, Ontario P3E 2C6, Canada}}
\newcommand{\Maryland}{\affiliation{Physics Department, University of Maryland, College Park, Maryland 20742, USA}}
\newcommand{\McGill}{\affiliation{Physics Department, McGill University, Montreal, Quebec H3A 2T8, Canada}}
\newcommand{\Munich}{\affiliation{Technische Universit\"at M\"unchen, Physikdepartment and Excellence Cluster Universe, Garching 80805, Germany}}
\newcommand{\SDakota}{\affiliation{Physics Department, University of South Dakota, Vermillion, South Dakota 57069, USA}}
\newcommand{\Seoul}{\affiliation{Department of Physics, University of Seoul, Seoul, Korea}}
\newcommand{\SLAC}{\affiliation{SLAC National Accelerator Laboratory, Menlo Park, California 94025, USA}}
\newcommand{\Stanford}{\affiliation{Physics Department, Stanford University, Stanford, California 94305, USA}}
\newcommand{\Stony}{\affiliation{Department of Physics and Astronomy, Stony Brook University, SUNY, Stony Brook, New York 11794, USA}}
\newcommand{\TRIUMF}{\affiliation{TRIUMF, Vancouver, British Columbia V6T 2A3, Canada}}
\newcommand{\UMass}{\affiliation{Amherst Center for Fundamental Interactions and Physics Department, University of Massachusetts, Amherst, MA 01003, USA}}
\newcommand{\WIPP}{\affiliation{Waste Isolation Pilot Plant, Carlsbad, New Mexico 88220, USA}}
\author{J.B.~Albert}\Indiana
\author{D.J.~Auty}\altaffiliation{Now at University of Alberta, Edmonton, Alberta, Canada}\Alabama
\author{P.S.~Barbeau}\Duke
\author{D.~Beck}\Illinois
\author{V.~Belov}\ITEP
\author{M.~Breidenbach}\SLAC
\author{T.~Brunner}\McGill\TRIUMF
\author{A.~Burenkov}\ITEP
\author{G.F.~Cao}\IHEP
\author{C.~Chambers}\CSU
\author{J.~Chaves}\Stanford
\author{B.~Cleveland}\altaffiliation{Also SNOLAB, Sudbury ON, Canada}\Laurentian
\author{M.~Coon}\Illinois
\author{A.~Craycraft}\CSU
\author{T.~Daniels}\SLAC
\author{M.~Danilov}\ITEP
\author{S.J.~Daugherty}\Indiana
\author{J.~Davis}\SLAC
\author{S.~Delaquis}\Bern
\author{A.~Der Mesrobian-Kabakian}\Laurentian
\author{R.~DeVoe}\Stanford
\author{T.~Didberidze}\Alabama
\author{J.~Dilling}\TRIUMF
\author{A.~Dolgolenko}\ITEP
\author{M.J.~Dolinski}\Drexel
\author{M.~Dunford}\Carleton
\author{W.~Fairbank Jr.}\CSU
\author{J.~Farine}\Laurentian
\author{W.~Feldmeier}\Munich
\author{S.~Feyzbakhsh}\UMass 	
\author{P.~Fierlinger}\Munich
\author{D.~Fudenberg}\Stanford
\author{R.~Gornea}\Carleton\TRIUMF
\author{K.~Graham}\Carleton
\author{G.~Gratta}\Stanford
\author{C.~Hall}\Maryland
\author{M.~Hughes}\Alabama
\author{M.J.~Jewell}\Stanford
\author{A.~Johnson}\SLAC
\author{T.N.~Johnson}\Indiana
\author{S.~Johnston}\UMass
\author{A.~Karelin}\ITEP
\author{L.J.~Kaufman}\Indiana
\author{R.~Killick}\Carleton
\author{J.~King}\UMass
\author{T.~Koffas}\Carleton
\author{S.~Kravitz}\Stanford
\author{R.~Kr\"{u}cken}\TRIUMF
\author{A.~Kuchenkov}\ITEP
\author{K.S.~Kumar}\Stony
\author{D.S.~Leonard}\IBS
\author{C.~Licciardi}\Carleton
\author{Y.H.~Lin}\Drexel
\author{J.~Ling}\altaffiliation{Now at Sun Yat-Sen University, Guangzhou, China}\Illinois
\author{R.~MacLellan}\SDakota
\author{M.G.~Marino}\Munich
\author{B.~Mong}\Laurentian
\author{D.~Moore}\Stanford
\author{O.~Njoya}\Stony
\author{R.~Nelson}\WIPP
\author{A.~Odian}\SLAC
\author{I.~Ostrovskiy}\Stanford
\author{A.~Piepke}\Alabama
\author{A.~Pocar}\UMass
\author{C.Y.~Prescott}\SLAC
\author{F.~Reti\`{e}re}\TRIUMF
\author{P.C.~Rowson}\SLAC
\author{J.J.~Russell}\SLAC
\author{A.~Schubert}\Stanford
\author{D.~Sinclair}\Carleton\TRIUMF
\author{E.~Smith}\Drexel
\author{V.~Stekhanov}\ITEP
\author{M.~Tarka}\Stony
\author{T.~Tolba}\altaffiliation{Now at Institute of High Energy Physics, Beijing, China}\Bern
\author{R.~Tsang}\Alabama
\author{K.~Twelker}\Stanford
\author{P.~Vogel}\CALTECH
\author{J.-L.~Vuilleumier}\Bern
\author{A.~Waite}\SLAC
\author{J.~Walton}\Illinois
\author{T.~Walton}\CSU
\author{M.~Weber}\Stanford
\author{L.J.~Wen}\IHEP
\author{U.~Wichoski}\Laurentian
\author{T.A.~Winick}\Drexel
\author{J.~Wood}\WIPP
\author{Q.Y.~Xu}\Stanford
\author{L.~Yang}\Illinois
\author{Y.-R.~Yen}\altaffiliation{Corresponding author: yung-ruey.yen@drexel.edu}\Drexel
\author{O.Ya.~Zeldovich}\ITEP

\collaboration{EXO-200 Collaboration}

\date{\today}

\begin{abstract}
EXO-200 is a single phase liquid xenon detector designed to search for neutrinoless $\beta\beta$ decay of $^{136}$Xe to the ground state of $^{136}$Ba. We report here on a search for the two-neutrino $\beta\beta$ decay of $^{136}$Xe to the first $0^+$ excited state, $0^+_1$, of $^{136}$Ba based on a 100 kg$\cdot$yr exposure of $^{136}$Xe. Using a specialized analysis employing a machine learning algorithm, we obtain a 90\% CL half-life sensitivity of $1.7 \times 10^{24}$ yr. We find no statistically significant evidence for the \bbtn decay to the excited state resulting in a lower limit of $T^{2\nu}_{1/2}$ ($0^+ \rightarrow 0^+_1$) $>$ 6.9 $\times$ 10$^{23}$ yr at 90\% CL. This observed limit is consistent with the estimated half-life of 2.5$\times$10$^{25}$ yr.

\end{abstract}

\maketitle

\section{\label{sec:intro}Introduction}

Nuclear double-beta ($\beta\beta$) decay with the emission of two neutrinos, first considered by Goeppert-Mayer in 1935 \cite{gm}, is a second-order weak transition observed in a number of even-even nuclei. The two-neutrino decay mode ($2\nu \beta \beta$) to the ground state was directly observed in nine nuclei with half-lives in excess of 10$^{18}$ yr, with $^{136}$Xe being the longest at 2.2 $\times$10$^{21}$ yr \cite{PDG, bbtn_improved}. Decays with half-lives up to 10$^{24}$ yr have been observed using indirect radiochemical and geochemical methods~\cite{Barabash:2010}.

The Standard Model allows \bbtn decays to the first $0^+$ excited state, denoted hereafter as $0^+_1$, of the daughter nucleus if this state is energetically accessible. These decays are suppressed relative to their ground-state counterparts and are generally accompanied by the emission of de-excitation $\gamma$s, creating a signature that is distinct from typical $\gamma$ backgrounds and \bbtn decays to the ground state. 

Measurements of decays to excited states may provide additional constraints on the nuclear matrix elements (NMEs) relevant to $\beta\beta$ decay. Using the NME for the excited state decay in a ratio between it and the ground-state decay would allow any shared uncertainties in the NMEs for these transitions to be canceled. Better knowledge of these NMEs could lead to a more precise determination of the effective Majorana neutrino mass from \bbzn half-life measurements \cite{duerr}. Searches for decays to excited states may also test exotic theories of alternate $\beta\beta$ decay mechanisms.

The first investigation of \bbtn decay to excited states was performed by Fiorini $et al$. in 1977 with $^{76}$Ge \cite{first_excited}. The first dedicated search was the Milano experiment in 1982, also with $^{76}$Ge \cite{Bellotti:1982tn}, while the first positive signal was obtained in 1995 for the \bbtn decay mode of $^{100}$Mo to the 0$^+_1$ excited state of $^{100}$Ru \cite{Mo_excited1995}. This decay was then precisely measured by the NEMO-3 experiment where all $\beta$ and $\gamma$ tracks were observed \cite{Mo}. \bbtn decays to excited states have also been observed in $^{150}$Nd in 2004 \cite{Nd} and again in 2014 where the $\gamma$ coincidence was explicitly measured \cite{Kidd2014}. More recently, current \bbzn experiments such as Gerda \cite{gerda} and KamLAND-Zen \cite{kz} have searched for decays to excited states using $^{76}$Ge and $^{136}$Xe, respectively.

\begin{figure}
\centering
\includegraphics[width=0.49\textwidth]{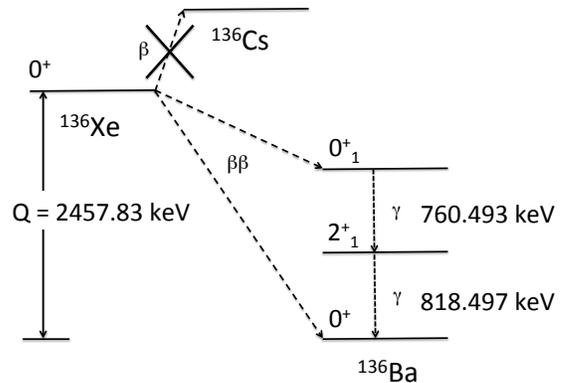}
\caption{Level scheme of the $\beta\beta$ decay of $^{136}$Xe. Decay to the excited state of $0^{+}_1$ will result in the emission of two $\gamma$s during the de-excitation to the ground state. The energy levels of the excited states are taken from \cite{nndc}. Only the $0^+$ and $2^+$ levels relevant for this search are shown.}\label{elevel_excited}
\end{figure}

Figure~\ref{elevel_excited} shows the energy level scheme of the $\beta\beta$ decay of $^{136}$Xe. A typical $\beta\beta$ decay with a $Q$ value of 878.8 keV transitions from the ground state of $^{136}$Xe to the $0^+_1$ state of $^{136}$Ba. The de-excitation from the $0^+_1$ state results in the emission of two $\gamma$s with energies of 760.5 keV and 818.5 keV. The two de-excitation $\gamma$s are emitted with a preferential angular correlation to be aligned or antialigned \cite{duerr}. Although the $2^+_1$ excited state of $^{136}$Ba is at lower energy than the $0^+_1$ state, the direct decay to it from the parent $0^+$ ground state of $^{136}$Xe is highly suppressed from angular momentum. The intermediate $2^+_1$ excited state has a half-life of 1.930 ps \cite{nndc}, which is not resolved temporally in most practical detectors. EXO-200 has the ability to identify both the $\beta\beta$ and de-excitation $\gamma $s in an excited state decay.

For decays to excited states, \bbtn decay to the $0^+_1$ state is expected to be the dominant decay mode, although \bbtn decay to the $2^+_1$ state or \bbzn decays to either $0^+_1$ or $2^+_1$ are also possible.  While EXO-200 can conduct searches for these other decays, this analysis focuses on the decay to the $0^+_1$ excited state. 

\section{\label{sec:theory}Theoretical Decay Rate}

The \bbtn decay rate for 0$^+$ $\rightarrow$ 0$^+$ transitions, including to the 0$^+_1$ excited state, can be written in the form,
\begin{equation}  \label{rate_bbtn}
[T^{2\nu}_{1/2}]^{-1} = G^{2\nu}(E_0,Z)\left| M^{2\nu}_{GT}-\frac{g^2_V}{g^2_A}M^{2\nu}_F \right| ^2,
\end{equation}
where $T^{2\nu}_{1/2}$ is the half-life, $M^{2\nu}_{GT}$ and $M^{2\nu}_{F}$ are the Gamow-Teller and Fermi nuclear transition matrix elements, respectively, $G^{2\nu}$ is the phase space integral that depends on the $Q$ value ($E_0$) and the atomic number of the daughter nucleus ($Z$), and $g_V$ and $g_A$ are the vector and axial-vector weak interaction coupling constants, respectively. The nuclear matrix elements (NMEs), $M^{2\nu}$, and phase space factor (PSF), $G^{2\nu}$, are expected to differ for decays to the ground and excited states. The PSF term results in a suppression of the decay rate because of the smaller $Q$ value for the transition to the excited state. 

The expected rate for the \bbtn decay of $^{136}$Xe to the first excited state is calculated from the measured \bbtn decay rate to the ground state and the theoretical values for the PSF and NMEs of both the ground and excited states. For the $^{136}$Xe \bbtn decays, the ratio of the PSF between the ground state and the excited state using the Schenter-Vogel Fermi function approximation and the Wilkinson correction for nuclear size is 3915 \cite{Vogel2012}. Using the PSF from~\cite{KotilaPhaseSpace}, a suppression ratio of 3956 is found. The two calculated values are within 1\% of each other, indicating that the theoretical uncertainty on the PSF suppression is small.

The NMEs for $\beta\beta$ decays have been calculated using the microscopic interacting boson model (IBM-2) with a method for isospin restoration and Argonne short-range correlations in the closure approximation in~\cite{Barea2015}. The ratio of the NMEs between the ground state and the excited state for $^{136}$Xe is 1.7. From Eq. (\ref{rate_bbtn}), this leads to an additional suppression factor of 2.9 for the decay rate to the excited state. The calculated NME values are model dependent and have larger theoretical uncertainty than the PSF. Including both suppression factors and applying them to the ground state half-life measured by EXO-200 of 2.16 $\times$ 10$^{21}$ yr \cite{bbtn_improved}, we estimate the expected \bbtn decay half-life of $^{136}$Xe to the 0$_1^+$ excited state of $^{136}$Ba of 2.5 $\times$ 10$^{25}$ yr. While this prediction puts the decay beyond the sensitivity of the present EXO-200 dataset, uncertainties in the NMEs may produce an overestimate of the half-life, as in the cases of $^{100}$Mo and $^{150}$Nd.

\section{\label{sec:detector}Detector Description}

The EXO-200 detector is a cylindrical, single-phase liquid xenon (LXe) time projection chamber (TPC), 40 cm in diameter and 44 cm in length, filled with xenon enriched to 80.6\% in $^{136}$Xe. Two drift regions are separated by a cathode at the center. A detailed description of the detector can be found in \cite{detector}. Radioimpurities in the detector components were minimized by a careful screening process \cite{radioassay} and a detector design optimized to use a minimal amount of materials. External radioactivity is reduced by $\geq$25 cm thick lead walls on all sides and additional passive shielding of $\geq$50 cm of high purity cryogenic fluid, HFE-7000 \cite{HFE}. The detector is located inside a clean room at the Waste Isolation Pilot Plant (WIPP) in Carlsbad, NM, USA, under an overburden of 1585$^{+11}_{-6}$ meters water equivalent~\cite{Esch:2005}. The remaining cosmic ray flux is detected by an active muon veto system consisting of plastic scintillation panels that cover the clean room on four sides. 

Energy depositions by ionizing radiation create free electrons and scintillation light in the LXe that are registered by anode wire grids and arrays of avalanche photodiodes, respectively. Two sets of wire grids form each anode. From the cathode to the anodes, charges will pass by V-wires (induction) first before being collected on the U-wires (charge collection). The U- and V- wire grids are offset by 60$^{\circ}$ to allow for two-dimensional (2D) reconstruction in the plane perpendicular to the axis of the TPC. The time difference between the prompt scintillation light and drifted charge collection allows for the position of the event in the drift direction ($Z$) to be determined.  Charge deposits (clusters) in a given event that are spatially separated by $\sim$1 cm or more can be individually resolved. An event can then be classified as single-site (SS), or multisite (MS), depending on the number of observed charge clusters. The total energy of an event is determined by combining the charge and scintillation signals, allowing improved energy resolution from the anticorrelation between these channels \cite{anticorrelation}. Radioactive $\gamma$ sources are periodically deployed at several positions near the TPC to characterize the detector response.

\section{\label{sec:method}Data Set and Methodology}

This search uses the same data set (``low-background data'')  and event selection criteria as the recent searches for \bbzn decay \cite{nature,majoron}. A total of 477.60$\pm$0.01 live days of data were collected between September 22, 2011 and September 1, 2013. Events consistent with noise, coincident with the muon veto, with more than one scintillation signal, or within $1$~s of other events in the TPC are removed. 

The fiducial volume is hexagonal with an apothem of 162 mm. Only regions within this hexagonal volume that are >10 mm from the cathode and anode wire planes are included.  This geometry corresponds to a $^{136}$Xe mass of 76.5 kg, or 3.39 $\times$ 10$^{26}$ atoms of $^{136}$Xe. The total exposure is 100 kg$\cdot$yr (736 mol$\cdot$yr). As in previous analyses, an energy range (summed over all charge clusters in an event) of 980$-$9800 keV is used. Finally, we require that all events have fully reconstructed U-, V-, and Z-positions.

A Geant4-based \cite{Geant4} Monte Carlo (MC) simulation of the detector and shielding (described in detail in \cite{bbtn_improved}) is used to model the detector response. The simulation of the \bbtn decay of $^{136}$Xe to the $0^+_1$ excited state of $^{136}$Ba (hereafter referred to simply as ``excited state events'') accounts for the smaller $Q$ value and angular correlation between the de-excitation $\gamma$s.  This MC is used to estimate the detection efficiency, determined by the percentage of these MC excited state events that survive all event selection cuts. The resulting efficiency for the excited state signal is (23.2~$\pm$~2.0)\%, with the dominant losses in efficiency arising from the fiducial volume and full reconstruction cuts. Errors in the estimate of this efficiency are accounted for in Sec.~\ref{sec:systematics}.

Based on periodic calibrations using $\gamma$ sources ($^{228}$Th, $^{60}$Co, $^{226}$Ra, and $^{137}$Cs), the energy scale and resolution are determined by fitting the full shape of the energy spectra observed in calibration data to MC simulations. In particular, the $^{60}$Co source produces events with multiple $\gamma$s of energies similar to those produced in the excited state decay. These calibration events show good agreement with MC simulations across all energies. The energy scale calibration and resolution are determined separately for SS and MS events.

To search for the \bbtn decay to the excited state, a binned maximum-likelihood (ML) fit is performed simultaneously over the SS and MS events using probability density functions (PDFs) in two dimensions: energy and an excited state ``discriminator'' variable. The PDFs are generated using the MC simulation and smeared with the appropriate energy resolution function. The same background components used in previous analyses \cite{nature,majoron} are also used here. The discriminator is specifically optimized to search for the unique event structure of the decay to the  $0^+_1$ excited state using machine learning techniques to be detailed in Sec.~\ref{sec:ML}. Using a 2D fit in energy and the machine learning discriminator improves the sensitivity to excited state events by more than a factor of three over the more generic technique in \cite{nature,majoron}, as detailed in Sec.~\ref{sec:sensitivity}.

\section{\label{sec:ML}Machine Learning}

This analysis uses machine learning techniques to create a variable (``discriminator'') which, for each event, indicates how ``signal-like'' (+1) or ``background-like'' (-1) it is. The machine learning software TMVA \cite{TMVA} (part of the ROOT data analysis framework \cite{ROOT}) is used to ``train'' an algorithm, using simulated signal and background data sets, to construct an optimized discriminator from a set of input variables and characterize the discriminator's performance.

\subsection{\label{sec:input}Input variables}
Prior to applying this algorithm to the low-background data, the input variables used to build the discriminator were finalized by optimizing the expected sensitivity to excited state events (as determined by the method described in Sec. \ref{sec:sensitivity}), while minimizing systematic errors resulting from disagreement between data and MC (discussed in Sec. \ref{sec:systematics}). By waiting to perform the fit until the method to determine the discriminator variable is finalized, potential biases from tuning the algorithm to the data set are minimized. The number of input variables is limited to reduce the sensitivity of the discriminator to systematic differences between the data and MC simulation. 

The majority of the discriminating power is provided by several event variables used in previous EXO-200 analyses \cite{nature,majoron}. These include the number of charge clusters in the event (multiplicity), the total event energy determined from the ionization and light response (energy), and the minimum separation of the charge deposits in the event from the anode wire plane or cylindrical walls of the TPC (standoff distance). The agreement between data and MC for these variables was studied in previous analyses \cite{bbtn_improved}. 

Although energy and multiplicity are included as inputs to the discriminator, the fit procedure also uses energy as the second fit dimension and separates the fit into SS and MS events. Including these additional dimensions allows the fit to better constrain individual background model components, while the discriminator is primarily useful for distinguishing between excited state events and all other backgrounds.

In addition, variables designed to take advantage of the energy deposits from the $^{136}$Ba de-excitation $\gamma$s are used. These variables, $\gamma_1$, $\gamma_2$, and $\gamma_{sum}$, are defined as:

\begin{equation}
{\rm \gamma_i }\equiv \underset{j\in S}{\rm min}	\left\{ \vert E_j - \epsilon_i \vert \right\},
\end{equation}
where $S$ is the set of charge clusters in an event, $E_j$ is the energy of cluster $j$, and $\epsilon_i $ are the de-excitation energies, $\epsilon_1 = 760.5$ keV, $\epsilon_2 = 818.5$ keV, and $\epsilon_{sum} = 1579$ keV. Hence, in the case where an event has a cluster of energy close to $\epsilon_i$, the corresponding variable $\gamma_i$ will be close to zero. Because only the total scintillation energy of an event is measured by the photodiodes, the $\gamma_i$ variables are determined using only the energy reconstructed from the charge signal of each cluster. Other variables incorporating charge cluster positions were investigated but showed no improvement in the method's sensitivity.

\subsection{Training data set}
\begin{figure*}[t]
    \centering	
		\includegraphics[width=0.99\textwidth]{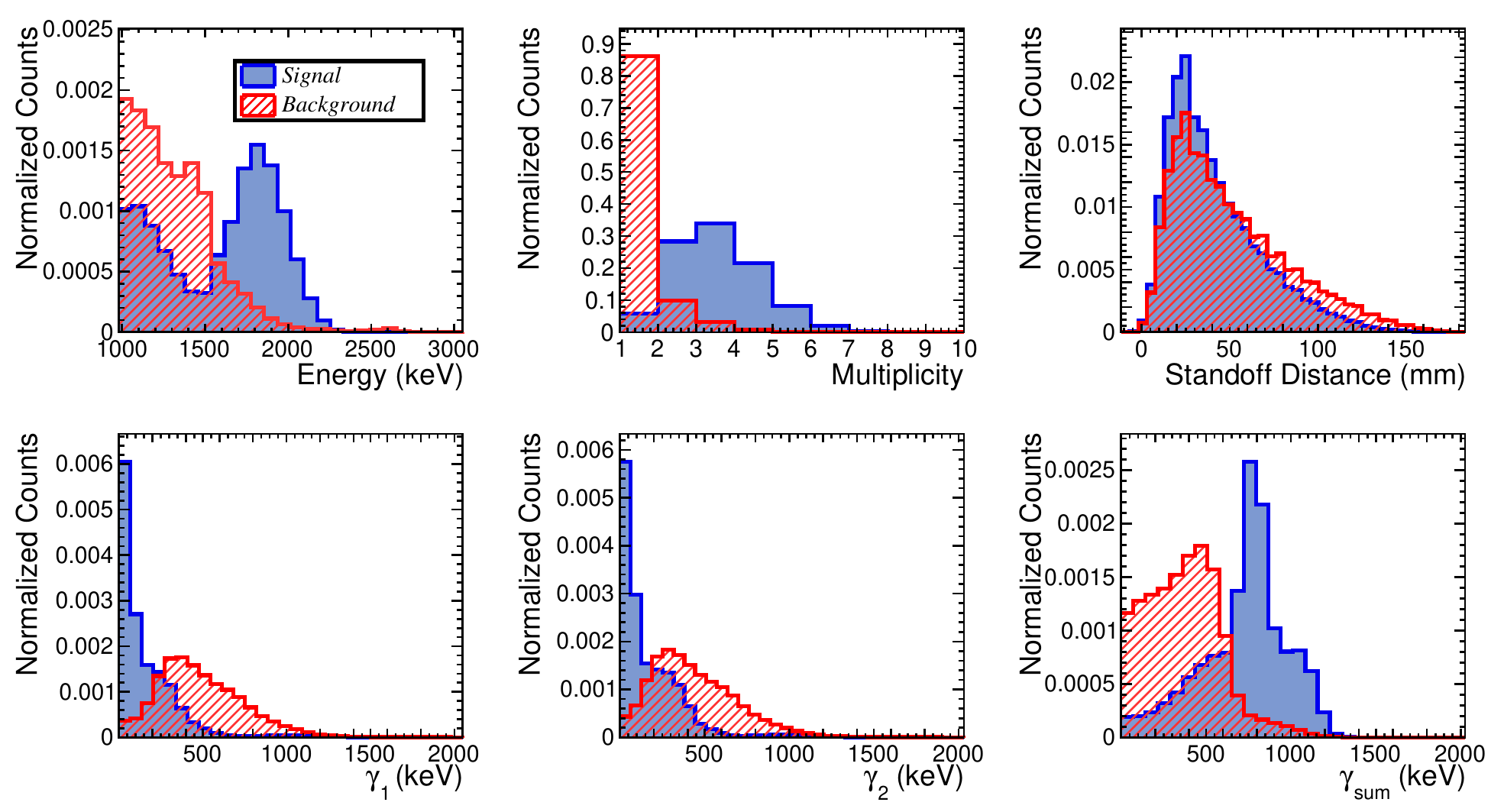}
	    \caption{(Color online) Distributions of the input variables to the discriminator based on MC simulations. Signal (excited state events) is shown in solid blue and background (all other events) in hatched red.}
	\label{fig:input_dists}
\end{figure*}

The training data consist of a signal class (excited state events) and a background class (all others). The dominant background to this search is \bbtn decays to the ground state, because they are the primary component of the low-background data. The background class was drawn from MC simulations, using a list of background components in quantities determined by the best fit model from a prior analysis of this data \cite{majoron}, to ensure that the training circumstances accurately reflect the low-background data set to which the discriminator will be applied. An equal number of excited state events were simulated to produce the signal class. All event selection cuts were applied prior to building this data set. In addition, both MC data classes were split in half, with one half used for training and the other for testing, to ensure that any patterns found in the training data set were not the result of statistical fluctuations. The distributions of each of the input variables used for this training data set are shown in Fig.~\ref{fig:input_dists}.

\subsection{Machine learning algorithm}
Several different machine learning algorithms were investigated, including boosted decision trees (BDTs), multilayer perceptrons (MLPs, a class of neural network), $k$-nearest neighbors (KNN), and simple rectangular cuts on the inputs. In preliminary tests of these algorithms using a cut on the discriminator variable to separate signal from background in the testing data set, the BDT provided the greatest discriminating power. Adjustments to the training parameters of the BDT did not indicate significant improvement by this metric. More detailed studies of the sensitivity and systematic errors of the three most promising of these algorithms (MLP and two BDT variants) were also performed, as described in Secs. \ref{sec:sensitivity} and \ref{sec:systematics}. These studies provided further evidence that the BDT algorithm is optimal among those considered and that changes to its training parameters have a small effect on its performance: The sensitivity for the MLP was roughly 50\% that of the BDT, and changing the BDT training parameters led to a 2\% increase in sensitivity at the cost of a 5\% increase in the systematic error. 

\subsection{Trained decision tree results} 
After training the BDT, the effectiveness of individual variables in deciding whether an event is signal-like or background-like can be determined. The BDT consists of many individual decision trees, each of which performs a series of binary cuts on the input variables, with the final nodes of the cuts each assigned to either signal (+1) or background (-1). The discriminator variable is then given by a sum of the individual trees' assignments, weighted by their classification performance. Further description of the BDT algorithm can be found in \cite{TMVA}. The ranking for any given variable is a measure of the fraction of decision tree cuts which use that variable, and is given in Table \ref{tab:var_ranking}. As expected, multiplicity is an effective discriminator between the decays to the excited state, which are largely MS, and \bbtn decays to the ground state, which are primarily SS, with most of the additional information contained in the energy variables.

\begin{table}[h]
\centering
\begin{tabular}{l c c}
\hline\hline 
 Rank & Variable &  Importance \\ 
 \hline
 1 & Multiplicity & 0.28 \\ 
 2 & Energy & 0.27 \\ 
 3 & $\gamma_{sum}$ & 0.14 \\ 
 4 & Standoff distance & 0.12 \\ 
 5 & $\gamma_1$ & 0.10 \\ 
 6 & $\gamma_2$ & 0.09 \\ 
 \hline\hline
 
\end{tabular}
\caption{Importance ranking of the input variables in the final boosted decision tree. The ``importance'' of each variable denotes the fraction of decision tree cuts which use that variable.}
\label{tab:var_ranking}
\end{table}

\begin{figure*}[t]
    \centering
		\includegraphics[width=0.49\textwidth]{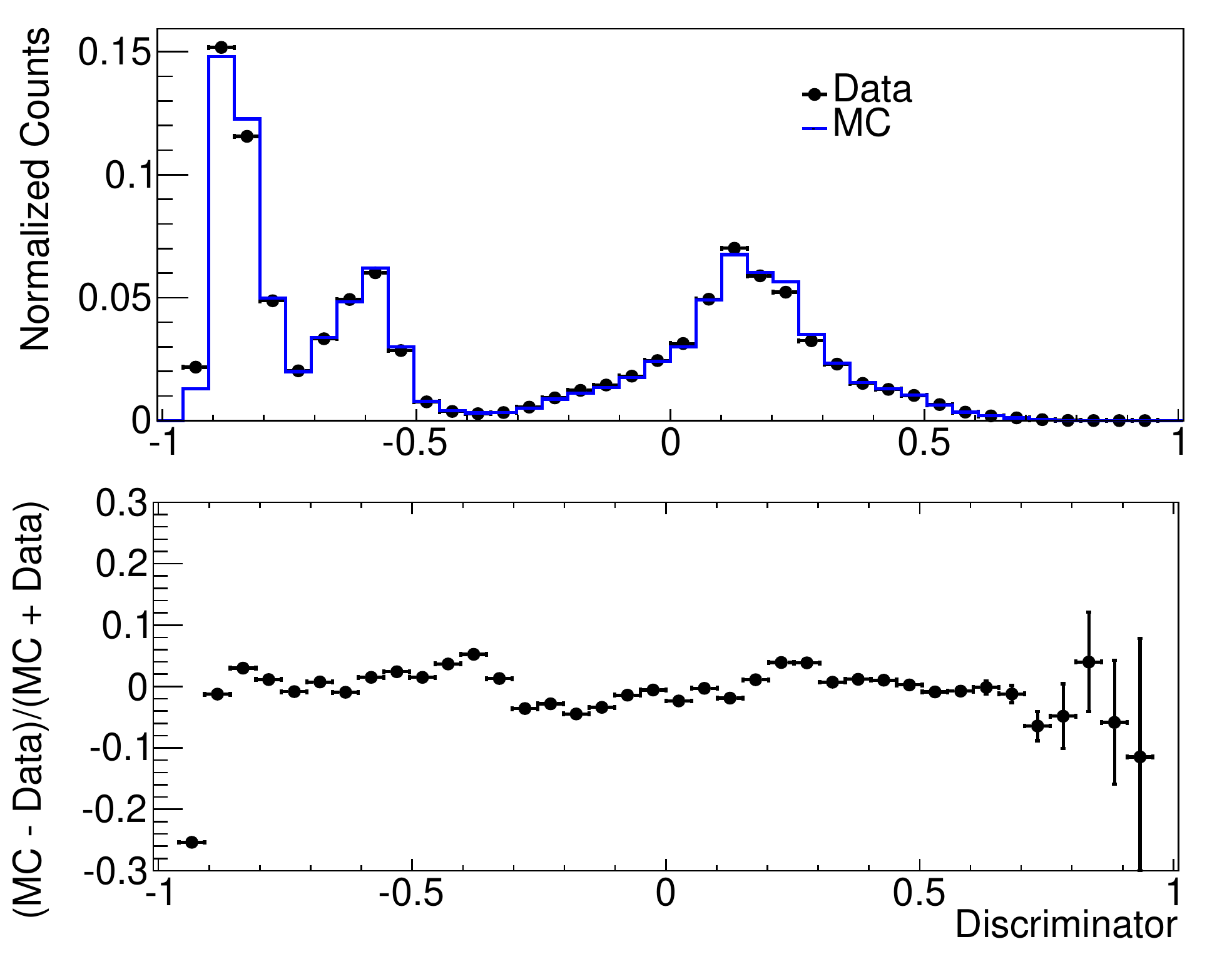}
		\includegraphics[width=0.49\textwidth]{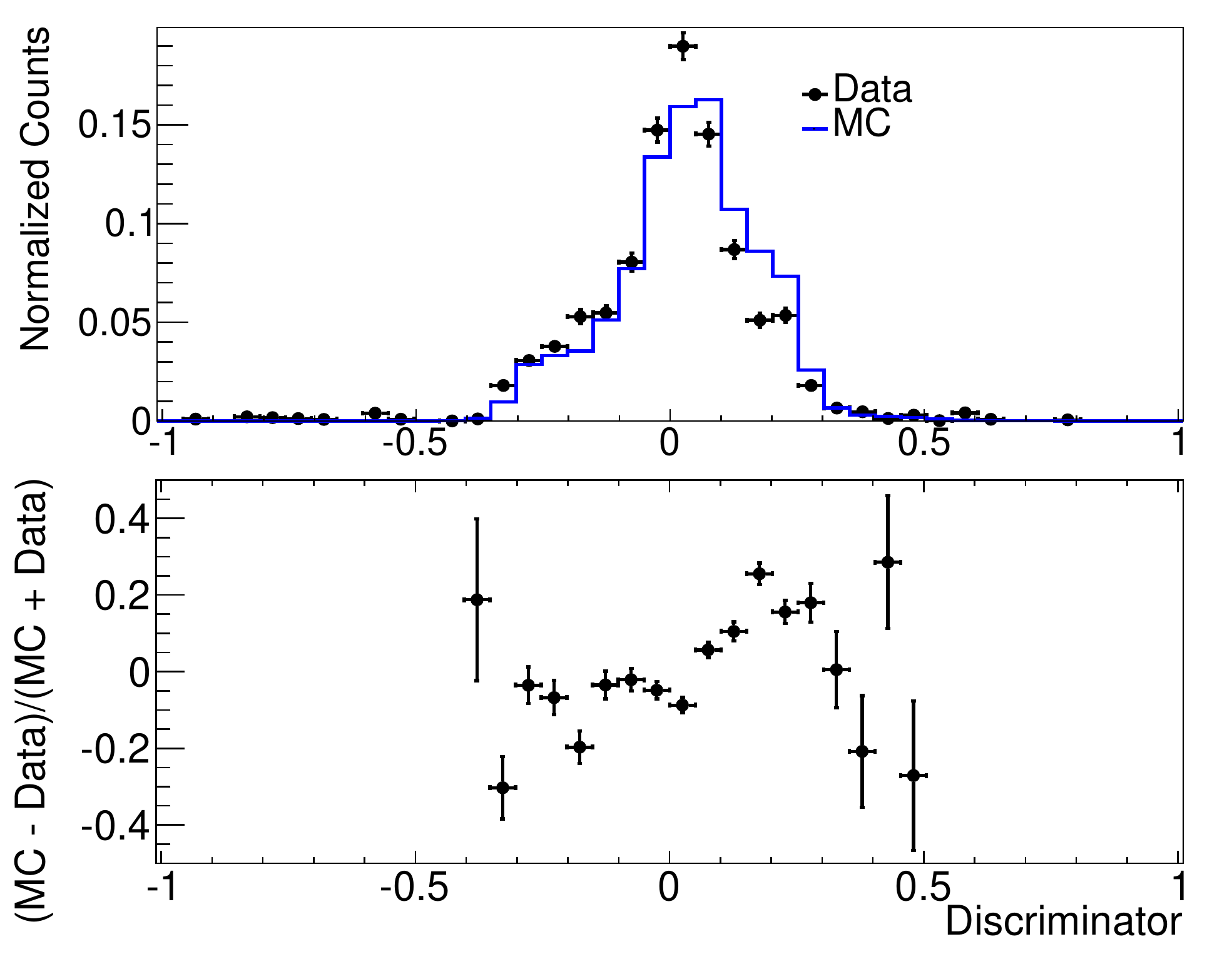}
	    \caption{Discriminator agreement between data and MC for MS $^{228}$Th (left) and MS \bbtn decays to the ground state (right). The \bbtn data are generated by subtraction of the non-\bbtn components, in amounts given by a prior analysis, from the low-background data. Residuals are shown in the bottom panels, ignoring those bins with insufficient counts.}
	\label{fig:source_agreement}
\end{figure*}

The trained BDT is then applied to low-background data, calibration data, and MC to determine the value of the discriminator variable for each event. In particular, this allows for comparison between data and MC for calibration sources, which can be used to quantify systematic effects. These comparisons are done for $^{60}$Co, $^{226}$Ra, and $^{228}$Th, both for SS and MS events. The primary $\gamma$ event backgrounds consist of MS events, for which a representative comparison using $^{228}$Th is shown in Fig.~\ref{fig:source_agreement}. While the data and MC distributions typically agree within 10\%, the remaining deviations are accounted for as systematic errors, to be discussed in Sec.~\ref{sec:systematics}.

Because \bbtn decays to the ground state constitute one of the largest backgrounds to this search, it is important to be able to quantify the difference in the shape of the discriminator variable distribution for data and MC for this component. To determine the \bbtn spectrum, each non-\bbtn component (from MC) is subtracted from the low-background data, in amounts given by a fit from a prior analysis \cite{majoron}. The final subtracted \bbtn distribution is then compared to the MC \bbtn component in the discriminator variable (Fig.~\ref{fig:source_agreement}). The results indicate good SS agreement, with some notable differences in the MS spectrum, which are accounted for in Sec. \ref{sec:systematics}. 

\subsection{\label{sec:sensitivity}Sensitivity and significance estimates}
Before the final fit to the data, the discriminator sensitivity was estimated by performing a two-dimensional profile likelihood fit on toy MC data sets. These MC data sets are created using the best-fit model from a previous analysis of the data \cite{majoron}, with no excited state events included. A fit in energy and discriminator is applied to find the 90\% confidence level (CL) upper limit on the number of excited state events, as determined by a profile of the negative log likelihood (NLL). An expected sensitivity of 43 events, corresponding to $T^{2\nu}_{1/2} = 1.7 \times 10^{24}$ yr, is given by the median upper limit obtained from toy MC data sets. This demonstrates an improvement by more than a factor of three over a simple fit to energy and standoff distance (as in \cite{nature,majoron}), which has a sensitivity of 170 events, corresponding to  $T^{2\nu}_{1/2} = 4.2 \times 10^{23}$ yr. 

The toy MC data sets with zero injected excited state events can also be used to calculate the significance of a nonzero final fit to data. The distribution of $\Delta$NLL values for fits with 0 excited state events is calculated; then, the fraction of toy MC data sets with $\Delta$NLL values greater than that found by the fit to data determines the level of compatibility with the null hypothesis. The full sensitivity and significance distributions are shown in Sec.~\ref{sec:results}.

\section{\label{sec:systematics}Systematic Uncertainties}

Systematic uncertainties in this analysis can be divided into those deriving from the event selection and background model, previously evaluated in \cite{nature}, and those unique to this analysis. In all cases, these uncertainties are accounted for by applying Gaussian constraints on the fit parameters. 

Unique to this analysis is an excited state event normalization term that accounts for discrepancies between MC and data in the discriminator and energy distributions of background components. To calculate this, toy MC data sets with a nonzero number of excited state events are generated from PDFs that have been skewed by the relative differences between data and MC, as measured in calibration source data (Fig.~\ref{fig:source_agreement}). These toy data sets are then fit to the standard, un-skewed PDFs, and the resulting difference between the fitted and simulated number of excited state events is determined. This fractional difference is accounted for as a systematic error which is applied in the normalization term specific to the excited state component.

While the deviations between calibration data and MC measured for the $^{228}$Th, $^{226}$Ra, and $^{60}$Co sources are used to skew the corresponding background components, the \bbtn ground state decay is skewed by deviations between MC and the background-subtracted data set. Differences between calibration sources are small, and the majority of the skewing error comes from the \bbtn component, so the exact choice of which calibration source is most similar to a given background component is inconsequential. 

\begin{figure*}[t!]
    \centering
		\includegraphics[width=0.8\textwidth]{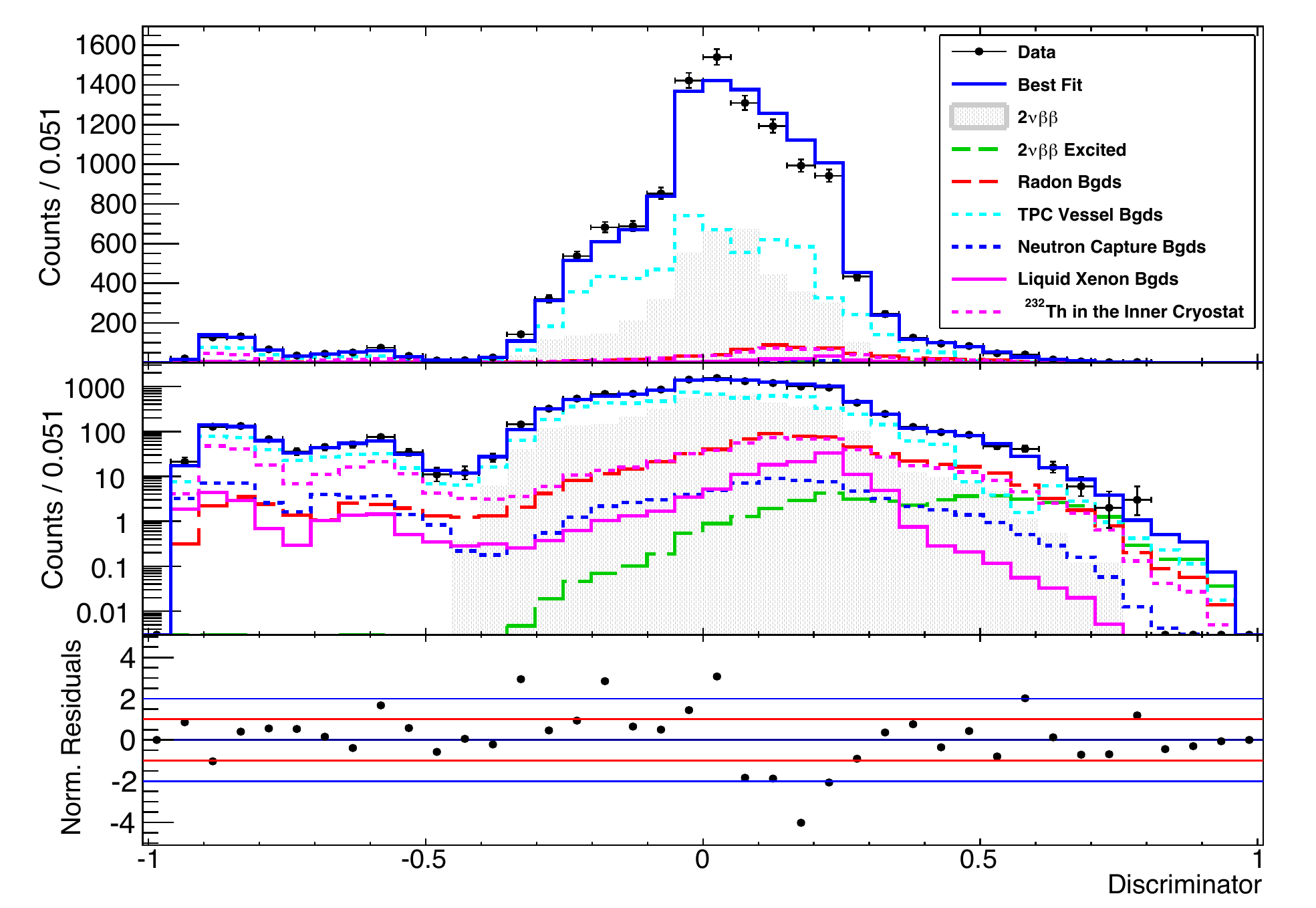}
	    \caption{(Color online) Fit results for the discriminator variable spectrum of MS events. The excited state event distribution is given by the dashed green line, concentrated toward positive discriminator values. Data points are shown in black and residuals between data and the best fit, normalized to the Poisson error, are presented, ignoring bins with 0 events.}
	\label{fig:final_fit_exc}
\end{figure*}

Because of statistical errors resulting from the background subtraction, the skewing cannot be precisely measured in the tail of the \bbtn distribution (\textit{i.e.}, values of the discriminator >0.5 or <-0.4 in the MS data). To prevent artificially distorting the distribution by errors resulting from the background subtraction, the skewing is forced to go to zero for bins with too few ($<20$) events. Several methods for suppressing the skewing in the low statistics tails were tested, with negligible effects on the resulting error. If an excited state decay signal were present in the data, its incomplete subtraction could lead to a slight overestimation of this systematic; however, while the estimation of this error may be conservative, this background-subtracted data set is only used for this study, so the procedure does not bias the final result. The final error found by this study is 15\%.

The additional Gaussian constraints are quantified in Table \ref{tab:uncertainties}. The SS fraction constraint is applied to the ratio of the number of SS events to the total number of events [SS/(SS+MS)] for each component, and is determined by the deviations in this ratio between data and MC for the calibration sources. A normalization constraint common to all components, accounting for the uncertainty in detection efficiency, is calculated from studies of the event reconstruction efficiency and the rate of events from calibration sources of known activity. A further normalization constraint is applied to account for uncertainties in the location of degenerate backgrounds. The relative fractions of neutron-capture related backgrounds, coming from cosmic ray muons and radioactive decays in the salt surrounding the experiment, are constrained according to MC studies and data coincident with muon-veto-panel events. The activity of radon in the liquid xenon is measured by observing characteristic $^{214}$Bi to $^{214}$Po coincidences in the detector, and the uncertainty on this measurement is translated into a constraint on the normalization of the radon components. 

The final systematic accounts for uncertainties in the ``$\beta$-scale'', which describes the possible difference in energy scales of $\beta$-like and $\gamma$-like events. Because the majority of the energy in the excited state events is deposited by the de-excitation $\gamma$s, these events are calibrated as $\gamma$-like components in the MC simulations. The $\beta$-scale variable is defined as an energy-independent ratio of $\gamma$ to $\beta$ energy scales and is allowed to float as a free parameter in the profile likelihood fit. We find it to be 0.9943 $\pm$ 0.0006.

Among the systematics, the SS fraction error had the largest effect on the final upper limit of the search. This uncertainty allows the largest background, \bbtn decays to the ground state, to shift from SS to MS, making them more signal-like.

\begin{table}[h]
\centering
\begin{tabular}{l c}
\hline\hline 
 Constraint & Error (\%) \\ 
 \hline
 Excited state normalization & 15 \\
 SS fraction & 4 \\
 Common normalization & 8.6 \\
 Background normalization & 20 \\
 Neutron capture fractions & 20 \\
 Radon in liquid xenon & 10 \\
 \hline\hline
 
\end{tabular}
\caption{Gaussian constraints applied to fit parameters to account for systematic uncertainties. These errors are explicitly included as input to the final fit to the low-background data, and are not the systematic uncertainties on the final result.}
\label{tab:uncertainties}
\end{table}

\begin{figure*}[t!]
   \centering
		\includegraphics[width=0.8\textwidth]{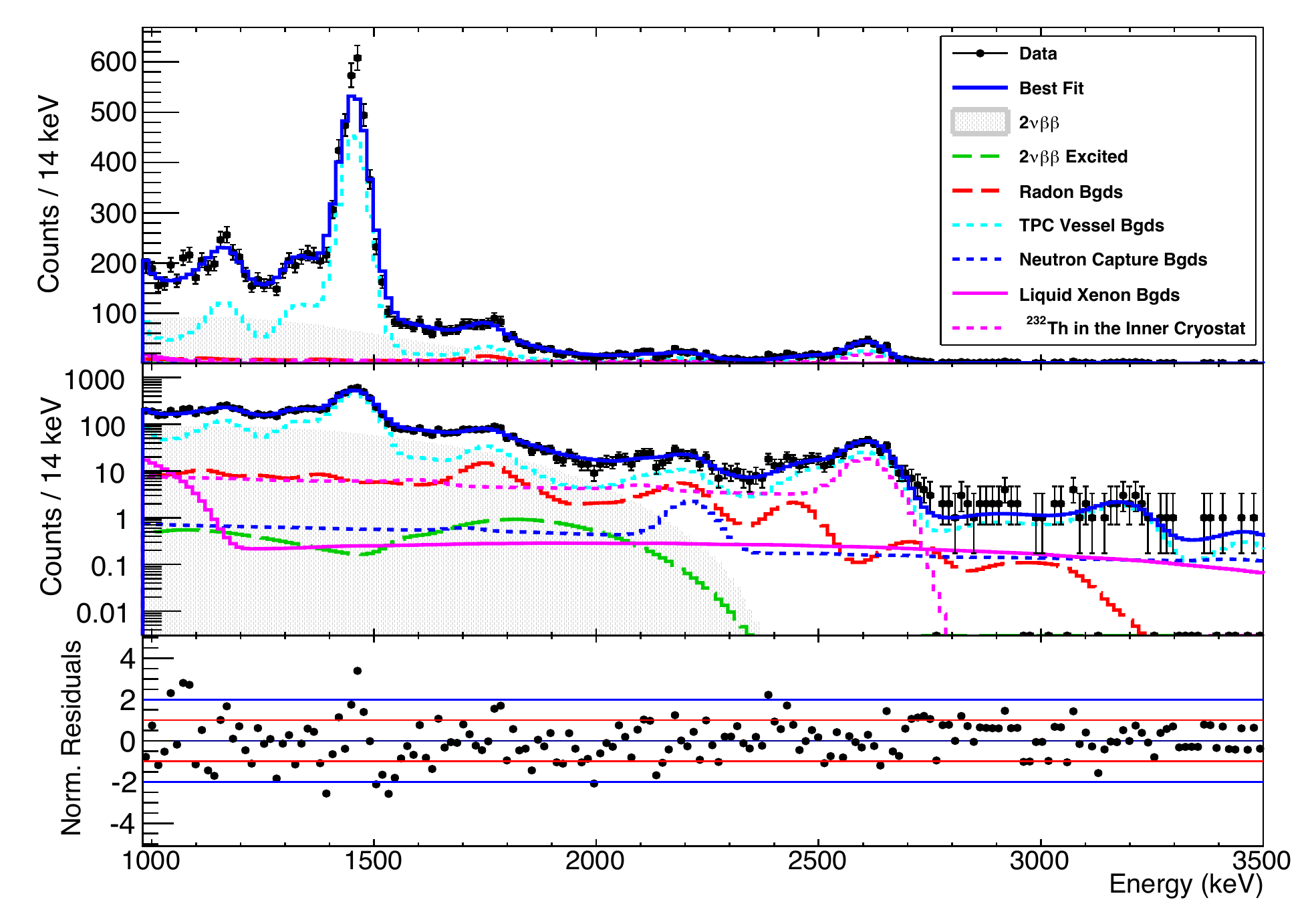}
	    \caption{(Color online) Fit results for the energy spectrum of MS events. Data points are shown in black and residuals between data and the best fit, normalized to the Poisson error, are presented, ignoring bins with 0 events.}
	\label{fig:final_fit_en}
\end{figure*}

\section{\label{sec:results}Results}

\begin{figure}[t!]
   \centering
		\includegraphics[width=0.49\textwidth]{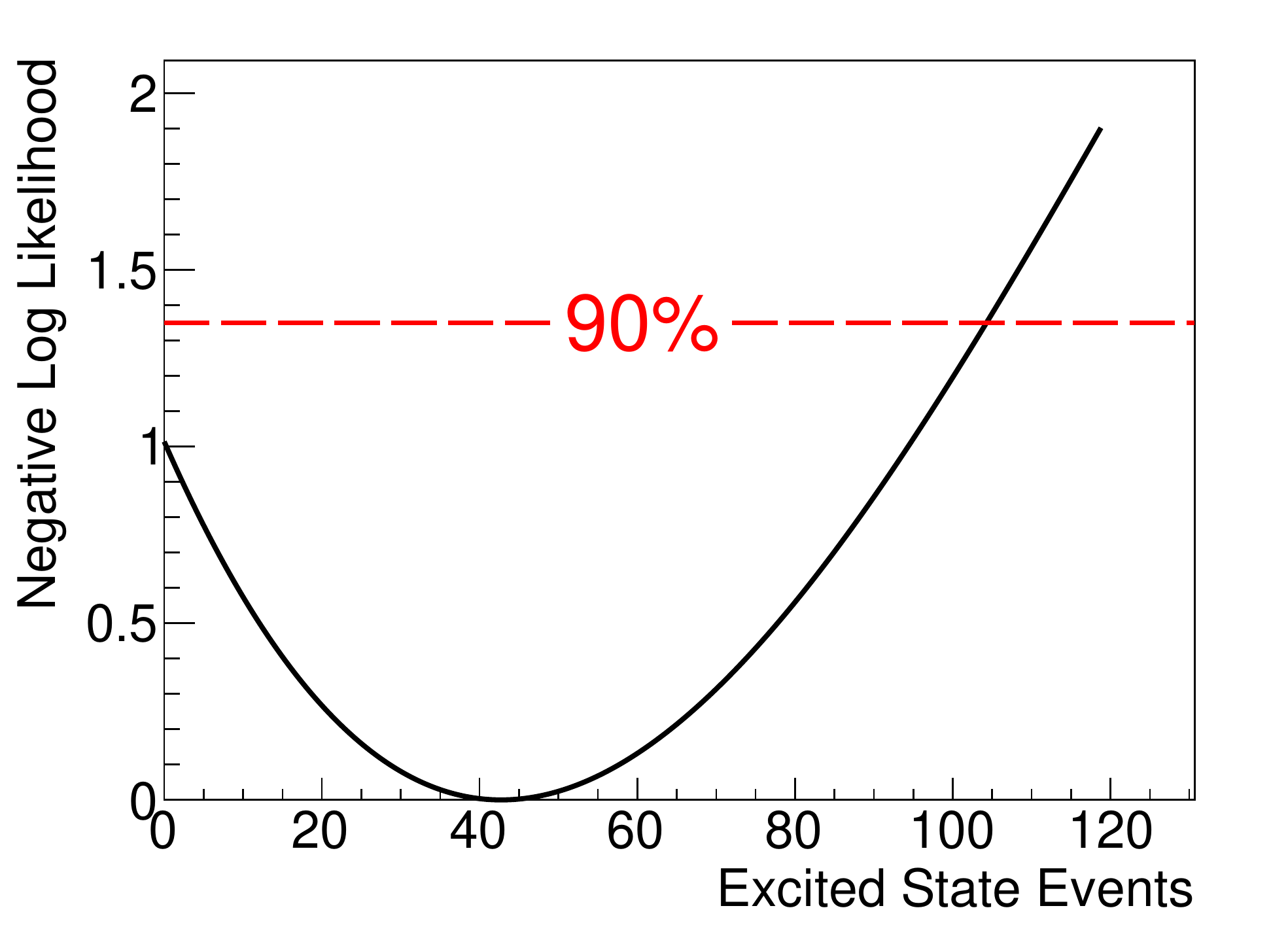}
	    \caption{NLL profile in the number of excited state events. The best fit value is 43 events, while the fit with 0 events has a $\Delta$NLL value of 1.0. The 90\% upper limit (dashed line) is 104 events.}
	\label{fig:prof_nexc}
\end{figure}

Plots of the discriminator variable and energy spectra of MS events for the fit to the data are shown in Figs.~\ref{fig:final_fit_exc} and \ref{fig:final_fit_en}. These plots illustrate the relative shapes and best fit quantities for all components, but do not contain all information used in the full fit, such as constraints on the SS fraction. The best fit values for background components are compatible with prior results, with the calculated half-life of the \bbtn decay to the ground state agreeing within systematic error \cite{bbtn_improved}. This fit was also checked for robustness against hypothetical backgrounds $^{88}$Y and $^{110m}$Ag, considered because of their mixed $\beta$ and $\gamma$ composition that could mimic an excited state signal. Both backgrounds were found to have a $<1 \sigma$ effect on the fit result, and prior analyses found no evidence of either component, so they are not included in the final result. The corresponding profile likelihood scan in the number of excited state events is shown in Fig.~\ref{fig:prof_nexc}. The profile finds a best-fit value of 43 events, with a 90\% CL upper limit of 104 events, assuming Wilks' theorem~\cite{Wilks:1938,Cowan} holds. Taking this result as a limit on the half-life of the decay to the $0_1^+$ excited state, we obtain $T^{2\nu}_{1/2} \geq 6.9 \times 10^{23}$ yr at 90\% CL. 

\begin{figure*}[t!]
    \centering
		\includegraphics[width=0.49\textwidth] {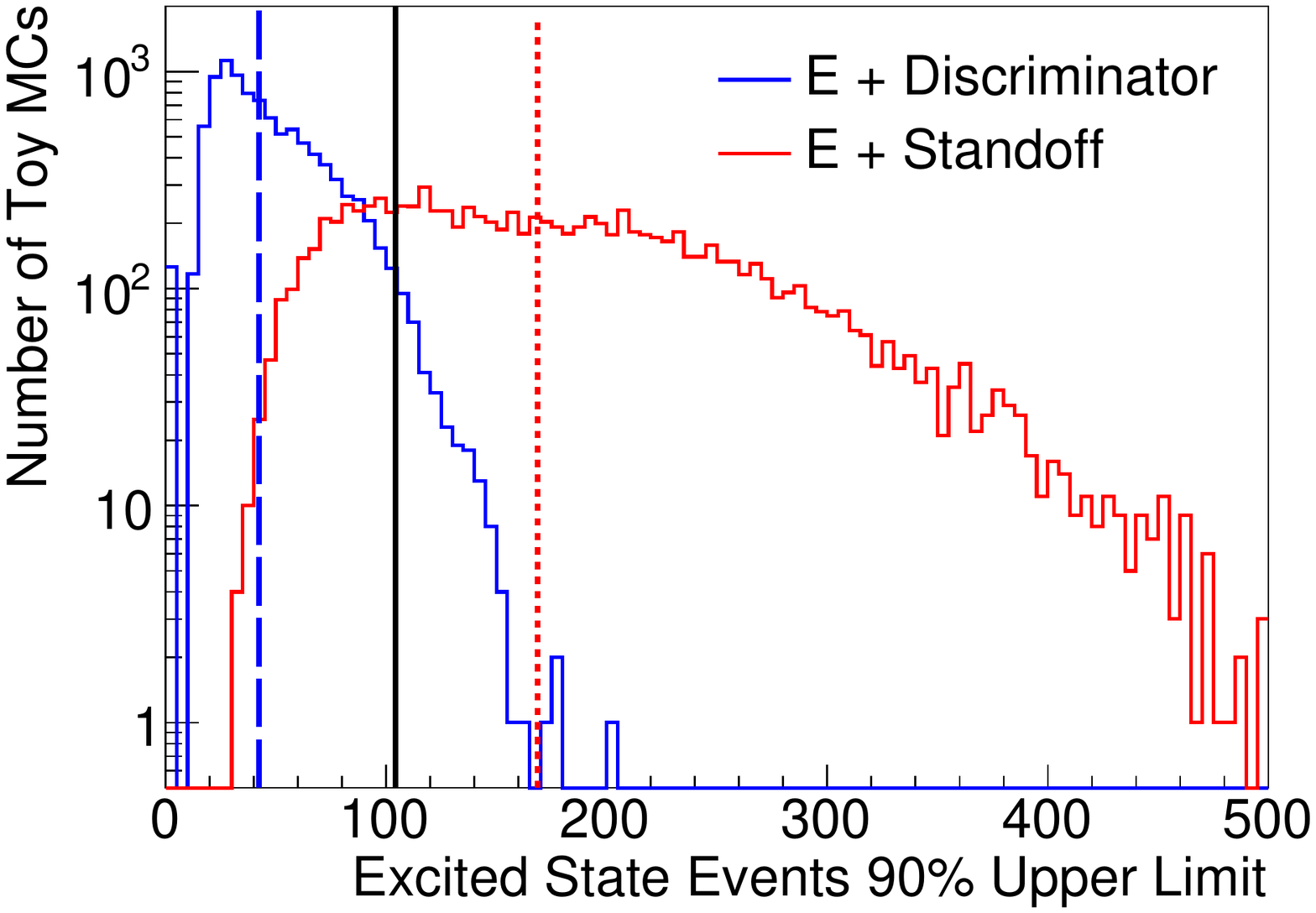}
		\includegraphics[width=0.49\textwidth] {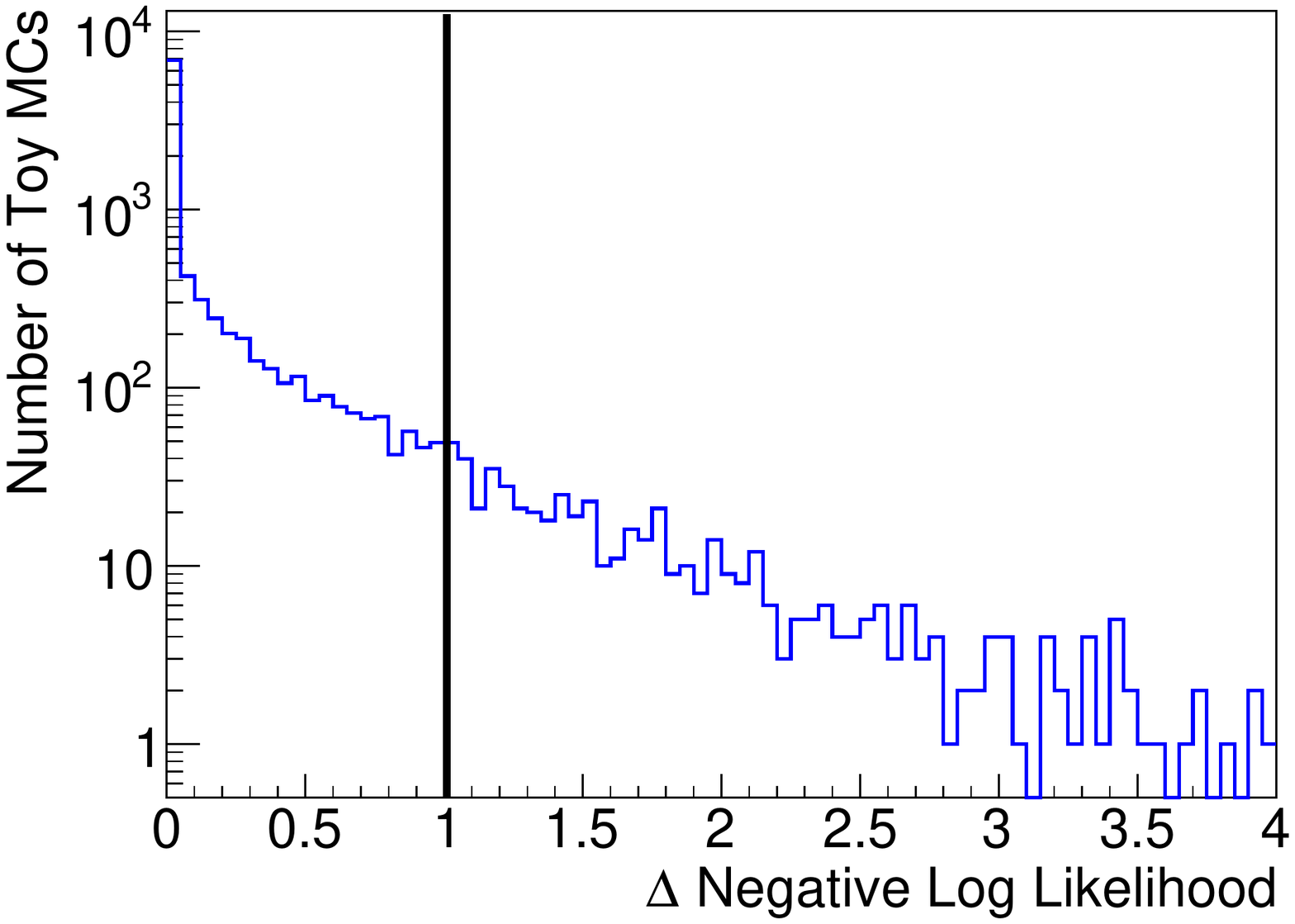}
	    \caption{(Color online) Left: Distribution of 90\% CL upper limits on the number of excited state events from toy MC data sets with no excited state events, using fits to energy and either standoff distance (red) or discriminator (blue). Median values are drawn with dashed lines, with the limit from data at 104 events drawn as a solid line. Right: Distribution of $\Delta$NLL values from toy MC data sets with no excited state events. The $\Delta$NLL for 0 events from data is shown as a solid line at 1.0.}
	\label{fig:significance}
\end{figure*}

The upper limit obtained from the data is roughly a factor of two weaker than the median expected sensitivity of $T^{2\nu}_{1/2} = 1.7 \times 10^{24}$ yr calculated in Sec. \ref{sec:sensitivity}. The calculation of significance, as described in Sec.~\ref{sec:sensitivity}, indicates that the result from data is consistent with the null hypothesis at 1.6$\sigma$ (Fig.~\ref{fig:significance}). Thus, this analysis does not find statistically significant evidence for a nonzero component of \bbtn decays to the excited state.

\section{\label{sec:summary}Conclusion}

We report here the results from a search for the \bbtn decay of $^{136}$Xe to the 0$_1^+$ excited state of $^{136}$Ba with the first two years of EXO-200 data, corresponding to a $^{136}$Xe exposure of 100 kg$\cdot$yr. No statistically significant evidence for this process is found. We obtain a limit on the half-life of this process of \mbox{$T^{2\nu}_{1/2}$ $>$ 6.9 $\times$ 10$^{23}$ yr} at 90\% CL, which is comparable to the recent results from KamLAND-Zen \cite{kz}. However, the ability of this analysis to identify the detailed event structure of excited state decays provides an expected sensitivity of \mbox{1.7 $\times$ 10$^{24}$ yr}, higher than the observed limit. This limit is consistent with theoretical calculations which predict, with substantial uncertainty, a half-life of $\sim 10^{25}$ yr; hence, sensitivity to this decay may be within reach of future analyses or a next-generation tonne-scale experiment.

\section{Acknowledgments}

The collaboration gratefully acknowledges Thilo Michel for valuable discussions, the KARMEN collaboration for supplying the cosmic-ray veto detectors, and the WIPP for their hospitality. EXO-200 is supported by DOE and NSF in the United States, NSERC in Canada, SNF in Switzerland, IBS in Korea, RFBR-14-02-00675 in Russia, DFG Cluster of Excellence ``Universe'' in Germany, and CAS-IHEP Fund and ISTCP (2015DFG02000) in China. EXO-200 data analysis and simulation uses resources of the National Energy Research Scientific Computing Center (NERSC), which is supported by the Office of Science of the U.S. Department of Energy under Contract No. DE-AC02-05CH11231.

\newpage
\bibliography{excited_ref_v3}

\begin{thebibliography}{29}%
\makeatletter
\providecommand \@ifxundefined [1]{%
 \@ifx{#1\undefined}
}%
\providecommand \@ifnum [1]{%
 \ifnum #1\expandafter \@firstoftwo
 \else \expandafter \@secondoftwo
 \fi
}%
\providecommand \@ifx [1]{%
 \ifx #1\expandafter \@firstoftwo
 \else \expandafter \@secondoftwo
 \fi
}%
\providecommand \natexlab [1]{#1}%
\providecommand \enquote  [1]{``#1''}%
\providecommand \bibnamefont  [1]{#1}%
\providecommand \bibfnamefont [1]{#1}%
\providecommand \citenamefont [1]{#1}%
\providecommand \href@noop [0]{\@secondoftwo}%
\providecommand \href [0]{\begingroup \@sanitize@url \@href}%
\providecommand \@href[1]{\@@startlink{#1}\@@href}%
\providecommand \@@href[1]{\endgroup#1\@@endlink}%
\providecommand \@sanitize@url [0]{\catcode `\\12\catcode `\$12\catcode
  `\&12\catcode `\#12\catcode `\^12\catcode `\_12\catcode `\%12\relax}%
\providecommand \@@startlink[1]{}%
\providecommand \@@endlink[0]{}%
\providecommand \url  [0]{\begingroup\@sanitize@url \@url }%
\providecommand \@url [1]{\endgroup\@href {#1}{\urlprefix }}%
\providecommand \urlprefix  [0]{URL }%
\providecommand \Eprint [0]{\href }%
\providecommand \doibase [0]{http://dx.doi.org/}%
\providecommand \selectlanguage [0]{\@gobble}%
\providecommand \bibinfo  [0]{\@secondoftwo}%
\providecommand \bibfield  [0]{\@secondoftwo}%
\providecommand \translation [1]{[#1]}%
\providecommand \BibitemOpen [0]{}%
\providecommand \bibitemStop [0]{}%
\providecommand \bibitemNoStop [0]{.\EOS\space}%
\providecommand \EOS [0]{\spacefactor3000\relax}%
\providecommand \BibitemShut  [1]{\csname bibitem#1\endcsname}%
\let\auto@bib@innerbib\@empty
\bibitem [{\citenamefont {{Goeppert-Mayer}}(1935)}]{gm}%
  \BibitemOpen
  \bibfield  {author} {\bibinfo {author} {\bibfnamefont {M.}~\bibnamefont
  {{Goeppert-Mayer}}},\ }\href {\doibase 10.1103/PhysRev.48.512} {\bibfield
  {journal} {\bibinfo  {journal} {Phys. Rev.}\ }\textbf {\bibinfo {volume}
  {48}},\ \bibinfo {pages} {512} (\bibinfo {year} {1935})}\BibitemShut
  {NoStop}%
\bibitem [{\citenamefont {Olive}\ \emph {et~al.}(2014)\citenamefont {Olive}
  \emph {et~al.}}]{PDG}%
  \BibitemOpen
  \bibfield  {author} {\bibinfo {author} {\bibfnamefont {K.~A.}\ \bibnamefont
  {Olive}} \emph {et~al.} (\bibinfo {collaboration} {Particle Data Group}),\
  }\href {\doibase 10.1088/1674-1137/38/9/090001} {\bibfield  {journal}
  {\bibinfo  {journal} {Chin. Phys.}\ }\textbf {\bibinfo {volume} {C38}},\
  \bibinfo {pages} {090001} (\bibinfo {year} {2014})}\BibitemShut {NoStop}%
\bibitem [{\citenamefont {Albert}\ \emph
  {et~al.}(2014{\natexlab{a}})\citenamefont {Albert} \emph
  {et~al.}}]{bbtn_improved}%
  \BibitemOpen
  \bibfield  {author} {\bibinfo {author} {\bibfnamefont {J.~B.}\ \bibnamefont
  {Albert}} \emph {et~al.} (\bibinfo {collaboration} {EXO-200}),\ }\href
  {\doibase 10.1103/PhysRevC.89.015502} {\bibfield  {journal} {\bibinfo
  {journal} {Phys. Rev. C}\ }\textbf {\bibinfo {volume} {89}},\ \bibinfo
  {pages} {015502} (\bibinfo {year} {2014}{\natexlab{a}})},\ \Eprint
  {http://arxiv.org/abs/1306.6106} {arXiv:1306.6106 [nucl-ex]} \BibitemShut
  {NoStop}%
\bibitem [{\citenamefont {Barabash}(2010)}]{Barabash:2010}%
  \BibitemOpen
  \bibfield  {author} {\bibinfo {author} {\bibfnamefont {A.~S.}\ \bibnamefont
  {Barabash}},\ }\href {\doibase 10.1103/PhysRevC.81.035501} {\bibfield
  {journal} {\bibinfo  {journal} {Phys. Rev. C}\ }\textbf {\bibinfo {volume}
  {81}},\ \bibinfo {pages} {035501} (\bibinfo {year} {2010})}\BibitemShut
  {NoStop}%
\bibitem [{\citenamefont {Duerr}\ \emph {et~al.}(2011)\citenamefont {Duerr},
  \citenamefont {Lindner},\ and\ \citenamefont {Zuber}}]{duerr}%
  \BibitemOpen
  \bibfield  {author} {\bibinfo {author} {\bibfnamefont {M.}~\bibnamefont
  {Duerr}}, \bibinfo {author} {\bibfnamefont {M.}~\bibnamefont {Lindner}}, \
  and\ \bibinfo {author} {\bibfnamefont {K.}~\bibnamefont {Zuber}},\ }\href
  {\doibase 10.1103/PhysRevD.84.093004} {\bibfield  {journal} {\bibinfo
  {journal} {Phys. Rev. D}\ }\textbf {\bibinfo {volume} {84}},\ \bibinfo
  {pages} {093004} (\bibinfo {year} {2011})},\ \Eprint
  {http://arxiv.org/abs/1103.4735} {arXiv:1103.4735 [hep-ph]} \BibitemShut
  {NoStop}%
\bibitem [{\citenamefont {Fiorini}(1978)}]{first_excited}%
  \BibitemOpen
  \bibfield  {author} {\bibinfo {author} {\bibfnamefont {E.}~\bibnamefont
  {Fiorini}},\ }\href@noop {} {\bibfield  {journal} {\bibinfo  {journal} {Proc.
  Int. Conf. NEUTRINO '77}\ }\textbf {\bibinfo {volume} {2}},\ \bibinfo {pages}
  {315} (\bibinfo {year} {1978})}\BibitemShut {NoStop}%
\bibitem [{\citenamefont {Bellotti}\ \emph {et~al.}(1982)\citenamefont
  {Bellotti}, \citenamefont {Fiorini}, \citenamefont {Liguori}, \citenamefont
  {Pullia}, \citenamefont {Sarracino},\ and\ \citenamefont
  {Zanotti}}]{Bellotti:1982tn}%
  \BibitemOpen
  \bibfield  {author} {\bibinfo {author} {\bibfnamefont {E.}~\bibnamefont
  {Bellotti}}, \bibinfo {author} {\bibfnamefont {E.}~\bibnamefont {Fiorini}},
  \bibinfo {author} {\bibfnamefont {C.}~\bibnamefont {Liguori}}, \bibinfo
  {author} {\bibfnamefont {A.}~\bibnamefont {Pullia}}, \bibinfo {author}
  {\bibfnamefont {A.}~\bibnamefont {Sarracino}}, \ and\ \bibinfo {author}
  {\bibfnamefont {L.}~\bibnamefont {Zanotti}},\ }\href {\doibase
  10.1007/BF02725993} {\bibfield  {journal} {\bibinfo  {journal} {Lett. Nuovo
  Cim.}\ }\textbf {\bibinfo {volume} {33}},\ \bibinfo {pages} {273} (\bibinfo
  {year} {1982})}\BibitemShut {NoStop}%
\bibitem [{\citenamefont {{Barabash}}\ \emph {et~al.}(1995)\citenamefont
  {{Barabash}} \emph {et~al.}}]{Mo_excited1995}%
  \BibitemOpen
  \bibfield  {author} {\bibinfo {author} {\bibfnamefont {A.~S.}\ \bibnamefont
  {{Barabash}}} \emph {et~al.},\ }\href {\doibase 10.1016/0370-2693(94)01657-X}
  {\bibfield  {journal} {\bibinfo  {journal} {Phys. Lett. B}\ }\textbf
  {\bibinfo {volume} {345}},\ \bibinfo {pages} {408} (\bibinfo {year}
  {1995})}\BibitemShut {NoStop}%
\bibitem [{\citenamefont {Kidd}\ \emph {et~al.}(2009)\citenamefont {Kidd},
  \citenamefont {Esterline}, \citenamefont {Tornow}, \citenamefont {Barabash},\
  and\ \citenamefont {Umatov}}]{Mo}%
  \BibitemOpen
  \bibfield  {author} {\bibinfo {author} {\bibfnamefont {M.~F.}\ \bibnamefont
  {Kidd}}, \bibinfo {author} {\bibfnamefont {J.~H.}\ \bibnamefont {Esterline}},
  \bibinfo {author} {\bibfnamefont {W.}~\bibnamefont {Tornow}}, \bibinfo
  {author} {\bibfnamefont {A.~S.}\ \bibnamefont {Barabash}}, \ and\ \bibinfo
  {author} {\bibfnamefont {V.~I.}\ \bibnamefont {Umatov}},\ }\href {\doibase
  10.1016/j.nuclphysa.2009.01.082} {\bibfield  {journal} {\bibinfo  {journal}
  {Nucl. Phys. A}\ }\textbf {\bibinfo {volume} {821}},\ \bibinfo {pages} {251}
  (\bibinfo {year} {2009})},\ \Eprint {http://arxiv.org/abs/0902.4418}
  {arXiv:0902.4418 [nucl-ex]} \BibitemShut {NoStop}%
\bibitem [{\citenamefont {Barabash}\ \emph {et~al.}(2004)\citenamefont
  {Barabash}, \citenamefont {Hubert}, \citenamefont {Hubert},\ and\
  \citenamefont {Umatov}}]{Nd}%
  \BibitemOpen
  \bibfield  {author} {\bibinfo {author} {\bibfnamefont {A.~S.}\ \bibnamefont
  {Barabash}}, \bibinfo {author} {\bibfnamefont {F.}~\bibnamefont {Hubert}},
  \bibinfo {author} {\bibfnamefont {P.}~\bibnamefont {Hubert}}, \ and\ \bibinfo
  {author} {\bibfnamefont {V.~I.}\ \bibnamefont {Umatov}},\ }\href {\doibase
  10.1134/1.1772463} {\bibfield  {journal} {\bibinfo  {journal} {Phys. Atom.
  Nucl.}\ }\textbf {\bibinfo {volume} {67}},\ \bibinfo {pages} {1216} (\bibinfo
  {year} {2004})}\BibitemShut {NoStop}%
\bibitem [{\citenamefont {Kidd}\ \emph {et~al.}(2014)\citenamefont {Kidd},
  \citenamefont {Esterline}, \citenamefont {Finch},\ and\ \citenamefont
  {Tornow}}]{Kidd2014}%
  \BibitemOpen
  \bibfield  {author} {\bibinfo {author} {\bibfnamefont {M.~F.}\ \bibnamefont
  {Kidd}}, \bibinfo {author} {\bibfnamefont {J.~H.}\ \bibnamefont {Esterline}},
  \bibinfo {author} {\bibfnamefont {S.~W.}\ \bibnamefont {Finch}}, \ and\
  \bibinfo {author} {\bibfnamefont {W.}~\bibnamefont {Tornow}},\ }\href
  {\doibase 10.1103/PhysRevC.90.055501} {\bibfield  {journal} {\bibinfo
  {journal} {Phys. Rev. C}\ }\textbf {\bibinfo {volume} {90}},\ \bibinfo
  {pages} {055501} (\bibinfo {year} {2014})},\ \Eprint
  {http://arxiv.org/abs/1411.3755} {arXiv:1411.3755 [nucl-ex]} \BibitemShut
  {NoStop}%
\bibitem [{\citenamefont {Agostini}\ \emph {et~al.}(2015)\citenamefont
  {Agostini} \emph {et~al.}}]{gerda}%
  \BibitemOpen
  \bibfield  {author} {\bibinfo {author} {\bibfnamefont {M.}~\bibnamefont
  {Agostini}} \emph {et~al.} (\bibinfo {collaboration} {GERDA}),\ }\href
  {\doibase 10.1088/0954-3899/42/11/115201} {\bibfield  {journal} {\bibinfo
  {journal} {J. Phys. G}\ }\textbf {\bibinfo {volume} {42}},\ \bibinfo {pages}
  {115201} (\bibinfo {year} {2015})},\ \Eprint
  {http://arxiv.org/abs/1506.03120} {arXiv:1506.03120 [hep-ex]} \BibitemShut
  {NoStop}%
\bibitem [{\citenamefont {Asakura}\ \emph {et~al.}(2015)\citenamefont {Asakura}
  \emph {et~al.}}]{kz}%
  \BibitemOpen
  \bibfield  {author} {\bibinfo {author} {\bibfnamefont {K.}~\bibnamefont
  {Asakura}} \emph {et~al.} (\bibinfo {collaboration} {KamLAND-Zen}),\
  }\href@noop {} {\  (\bibinfo {year} {2015})},\ \Eprint
  {http://arxiv.org/abs/1509.03724} {arXiv:1509.03724 [hep-ex]} \BibitemShut
  {NoStop}%
\bibitem [{\citenamefont {{Sonzogni}}(2002)}]{nndc}%
  \BibitemOpen
  \bibfield  {author} {\bibinfo {author} {\bibfnamefont {A.~A.}\ \bibnamefont
  {{Sonzogni}}},\ }\href@noop {} {\bibfield  {journal} {\bibinfo  {journal}
  {Nuclear Data Sheets}\ }\textbf {\bibinfo {volume} {95}},\ \bibinfo {pages}
  {837} (\bibinfo {year} {2002})},\ \bibinfo {note} {data extracted from the
  ENSDF database, Oct. 9, 2015, http://www.nndc.bnl.gov/}\BibitemShut {NoStop}%
\bibitem [{\citenamefont {Vogel}(2012)}]{Vogel2012}%
  \BibitemOpen
  \bibfield  {author} {\bibinfo {author} {\bibfnamefont {P.}~\bibnamefont
  {Vogel}},\ }\href@noop {} {}\bibinfo {howpublished} {private communication}
  (\bibinfo {year} {2012})\BibitemShut {NoStop}%
\bibitem [{\citenamefont {Kotila}\ and\ \citenamefont
  {Iachello}(2012)}]{KotilaPhaseSpace}%
  \BibitemOpen
  \bibfield  {author} {\bibinfo {author} {\bibfnamefont {J.}~\bibnamefont
  {Kotila}}\ and\ \bibinfo {author} {\bibfnamefont {F.}~\bibnamefont
  {Iachello}},\ }\href {\doibase 10.1103/PhysRevC.85.034316} {\bibfield
  {journal} {\bibinfo  {journal} {Phys. Rev. C}\ }\textbf {\bibinfo {volume}
  {85}},\ \bibinfo {pages} {034316} (\bibinfo {year} {2012})},\ \Eprint
  {http://arxiv.org/abs/1209.5722} {arXiv:1209.5722 [nucl-th]} \BibitemShut
  {NoStop}%
\bibitem [{\citenamefont {Barea}\ \emph {et~al.}(2015)\citenamefont {Barea},
  \citenamefont {Kotila},\ and\ \citenamefont {Iachello}}]{Barea2015}%
  \BibitemOpen
  \bibfield  {author} {\bibinfo {author} {\bibfnamefont {J.}~\bibnamefont
  {Barea}}, \bibinfo {author} {\bibfnamefont {J.}~\bibnamefont {Kotila}}, \
  and\ \bibinfo {author} {\bibfnamefont {F.}~\bibnamefont {Iachello}},\ }\href
  {\doibase 10.1103/PhysRevC.91.034304} {\bibfield  {journal} {\bibinfo
  {journal} {Phys. Rev. C}\ }\textbf {\bibinfo {volume} {91}},\ \bibinfo
  {pages} {034304} (\bibinfo {year} {2015})},\ \Eprint
  {http://arxiv.org/abs/1506.08530} {arXiv:1506.08530 [nucl-th]} \BibitemShut
  {NoStop}%
\bibitem [{\citenamefont {Auger}\ \emph {et~al.}(2012)\citenamefont {Auger}
  \emph {et~al.}}]{detector}%
  \BibitemOpen
  \bibfield  {author} {\bibinfo {author} {\bibfnamefont {M.}~\bibnamefont
  {Auger}} \emph {et~al.},\ }\href {\doibase 10.1088/1748-0221/7/05/P05010}
  {\bibfield  {journal} {\bibinfo  {journal} {JINST}\ }\textbf {\bibinfo
  {volume} {7}},\ \bibinfo {pages} {P05010} (\bibinfo {year} {2012})},\ \Eprint
  {http://arxiv.org/abs/1202.2192} {arXiv:1202.2192 [physics.ins-det]}
  \BibitemShut {NoStop}%
\bibitem [{\citenamefont {Leonard}\ \emph {et~al.}(2008)\citenamefont {Leonard}
  \emph {et~al.}}]{radioassay}%
  \BibitemOpen
  \bibfield  {author} {\bibinfo {author} {\bibfnamefont {D.~S.}\ \bibnamefont
  {Leonard}} \emph {et~al.},\ }\href {\doibase 10.1016/j.nima.2008.03.001}
  {\bibfield  {journal} {\bibinfo  {journal} {Nucl. Instrum. Meth. A}\ }\textbf
  {\bibinfo {volume} {591}},\ \bibinfo {pages} {490} (\bibinfo {year}
  {2008})},\ \Eprint {http://arxiv.org/abs/0709.4524} {arXiv:0709.4524
  [physics.ins-det]} \BibitemShut {NoStop}%
\bibitem [{HFE()}]{HFE}%
  \BibitemOpen
  \href@noop {} {\enquote {\bibinfo {title} {{3M HFE-7000}},}\ }\bibinfo
  {howpublished} {\url{http://www.3m.com/}}\BibitemShut {NoStop}%
\bibitem [{\citenamefont {{Esch}}\ \emph {et~al.}(2005)\citenamefont {{Esch}},
  \citenamefont {{Bowles}}, \citenamefont {{Hime}}, \citenamefont
  {{Pichlmaier}}, \citenamefont {{Reifarth}},\ and\ \citenamefont
  {{Wollnik}}}]{Esch:2005}%
  \BibitemOpen
  \bibfield  {author} {\bibinfo {author} {\bibfnamefont {E.-I.}\ \bibnamefont
  {{Esch}}}, \bibinfo {author} {\bibfnamefont {T.~J.}\ \bibnamefont
  {{Bowles}}}, \bibinfo {author} {\bibfnamefont {A.}~\bibnamefont {{Hime}}},
  \bibinfo {author} {\bibfnamefont {A.}~\bibnamefont {{Pichlmaier}}}, \bibinfo
  {author} {\bibfnamefont {R.}~\bibnamefont {{Reifarth}}}, \ and\ \bibinfo
  {author} {\bibfnamefont {H.}~\bibnamefont {{Wollnik}}},\ }\href {\doibase
  10.1016/j.nima.2004.09.005} {\bibfield  {journal} {\bibinfo  {journal}
  {Nuclear Instruments and Methods in Physics Research A}\ }\textbf {\bibinfo
  {volume} {538}},\ \bibinfo {pages} {516} (\bibinfo {year} {2005})},\ \Eprint
  {http://arxiv.org/abs/astro-ph/0408486} {astro-ph/0408486} \BibitemShut
  {NoStop}%
\bibitem [{\citenamefont {Conti}\ \emph {et~al.}(2003)\citenamefont {Conti}
  \emph {et~al.}}]{anticorrelation}%
  \BibitemOpen
  \bibfield  {author} {\bibinfo {author} {\bibfnamefont {E.}~\bibnamefont
  {Conti}} \emph {et~al.},\ }\href {\doibase 10.1103/PhysRevB.68.054201}
  {\bibfield  {journal} {\bibinfo  {journal} {Phys. Rev. B}\ }\textbf {\bibinfo
  {volume} {68}},\ \bibinfo {pages} {054201} (\bibinfo {year} {2003})},\
  \Eprint {http://arxiv.org/abs/hep-ex/0303008} {arXiv:hep-ex/0303008 [hep-ex]}
  \BibitemShut {NoStop}%
\bibitem [{\citenamefont {Albert}\ \emph
  {et~al.}(2014{\natexlab{b}})\citenamefont {Albert} \emph {et~al.}}]{nature}%
  \BibitemOpen
  \bibfield  {author} {\bibinfo {author} {\bibfnamefont {J.~B.}\ \bibnamefont
  {Albert}} \emph {et~al.} (\bibinfo {collaboration} {EXO-200}),\ }\href
  {\doibase 10.1038/nature13432} {\bibfield  {journal} {\bibinfo  {journal}
  {Nature}\ }\textbf {\bibinfo {volume} {510}},\ \bibinfo {pages} {229}
  (\bibinfo {year} {2014}{\natexlab{b}})},\ \Eprint
  {http://arxiv.org/abs/1402.6956} {arXiv:1402.6956 [nucl-ex]} \BibitemShut
  {NoStop}%
\bibitem [{\citenamefont {Albert}\ \emph
  {et~al.}(2014{\natexlab{c}})\citenamefont {Albert} \emph {et~al.}}]{majoron}%
  \BibitemOpen
  \bibfield  {author} {\bibinfo {author} {\bibfnamefont {J.~B.}\ \bibnamefont
  {Albert}} \emph {et~al.} (\bibinfo {collaboration} {EXO-200}),\ }\href
  {\doibase 10.1103/PhysRevD.90.092004} {\bibfield  {journal} {\bibinfo
  {journal} {Phys. Rev. D}\ }\textbf {\bibinfo {volume} {90}},\ \bibinfo
  {pages} {092004} (\bibinfo {year} {2014}{\natexlab{c}})},\ \Eprint
  {http://arxiv.org/abs/1409.6829} {arXiv:1409.6829 [hep-ex]} \BibitemShut
  {NoStop}%
\bibitem [{\citenamefont {Allison}\ \emph {et~al.}(2006)\citenamefont {Allison}
  \emph {et~al.}}]{Geant4}%
  \BibitemOpen
  \bibfield  {author} {\bibinfo {author} {\bibfnamefont {J.}~\bibnamefont
  {Allison}} \emph {et~al.},\ }\href {\doibase 10.1109/TNS.2006.869826}
  {\bibfield  {journal} {\bibinfo  {journal} {IEEE Trans. Nucl. Sci.}\ }\textbf
  {\bibinfo {volume} {53}},\ \bibinfo {pages} {270} (\bibinfo {year}
  {2006})}\BibitemShut {NoStop}%
\bibitem [{\citenamefont {Hocker}\ \emph {et~al.}(2007)\citenamefont {Hocker}
  \emph {et~al.}}]{TMVA}%
  \BibitemOpen
  \bibfield  {author} {\bibinfo {author} {\bibfnamefont {A.}~\bibnamefont
  {Hocker}} \emph {et~al.},\ }\href@noop {} {\bibfield  {journal} {\bibinfo
  {journal} {PoS (ACAT 2007)}\ }\textbf {\bibinfo {volume} {ACAT}},\ \bibinfo
  {pages} {040} (\bibinfo {year} {2007})},\ \Eprint
  {http://arxiv.org/abs/physics/0703039} {arXiv:physics/0703039 [physics]}
  \BibitemShut {NoStop}%
\bibitem [{\citenamefont {Brun}\ and\ \citenamefont {Rademakers}(1997)}]{ROOT}%
  \BibitemOpen
  \bibfield  {author} {\bibinfo {author} {\bibfnamefont {R.}~\bibnamefont
  {Brun}}\ and\ \bibinfo {author} {\bibfnamefont {F.}~\bibnamefont
  {Rademakers}},\ }\href {\doibase 10.1016/S0168-9002(97)00048-X} {\bibfield
  {journal} {\bibinfo  {journal} {Nucl. Instrum. Meth. A}\ }\textbf {\bibinfo
  {volume} {389}},\ \bibinfo {pages} {81} (\bibinfo {year} {1997})}\BibitemShut
  {NoStop}%
\bibitem [{\citenamefont {Wilks}(1938)}]{Wilks:1938}%
  \BibitemOpen
  \bibfield  {author} {\bibinfo {author} {\bibfnamefont {S.~S.}\ \bibnamefont
  {Wilks}},\ }\href {\doibase 10.1214/aoms/1177732360} {\bibfield  {journal}
  {\bibinfo  {journal} {Annals Math. Statist.}\ }\textbf {\bibinfo {volume}
  {9}},\ \bibinfo {pages} {60} (\bibinfo {year} {1938})}\BibitemShut {NoStop}%
\bibitem [{\citenamefont {Cowan}(1998)}]{Cowan}%
  \BibitemOpen
  \bibfield  {author} {\bibinfo {author} {\bibfnamefont {G.}~\bibnamefont
  {Cowan}},\ }\href@noop {} {\emph {\bibinfo {title} {Statistical Data
  Analysis}}}\ (\bibinfo  {publisher} {Oxford Science Publications,
  Clarendon},\ \bibinfo {year} {1998})\BibitemShut {NoStop}%
\end{thebibliography}%

\end{document}